\documentclass[pra, 12 pt]{revtex4}
\usepackage[english]{babel}
\usepackage{amsmath}
\usepackage{epsfig}

\begin{document}

\title{Remarks on the Nyquist and Callen-Welton Theorems}
\author{S.A. Trigger $^{1}$,\, G.J.F. van Heijst $^{2}$,\,
A.A. Rukhadze $^{3}$,\, and P.P.J.M. Schram $^{2}$\,}
\address{{1} Joint\, Institute\, for\, High\,
Temperatures, Russian\, Academy\, of\, Sciences, 13/19, Izhorskaia
Str., Moscow\, 127412, Russia; email:\,strig@gmx.net}
\address{{2} Eindhoven\, University\, of\, Technology, P.O.\, Box\, 513,
MB\, 5600\, Eindhoven, The\, Netherlands}
\address{{3} A.M. Prokhorov\,
General\, Physics\, Institute\, Moscow\,, 119991, Russia}

\begin{abstract}

The connection of the Callen-Welton and Nyquist
fluctuation-dissipation relations is considered for plasma-like
classical and quantum systems. The conditions for appearance of
the dissipative parameters in the equilibrium current-current
correlation function are investigated. The paper presents the
arguments for the restrictions of the Nyquist theorem and against
violation of the Callen-Welton theorem in the quantum case.

\end{abstract}

\maketitle

\section{Introduction}

The Callen-Welton theorem (CWT) is a general and rigorous result
of statistical physics (see e.g. \cite{LL1}). Some ideas have been
suggested \cite{KL1, VSch04} to show that the existing form of CWT
in the quantum case is not correct. One of the arguments is based
on a consideration of the oscillatory electric circuit. As is well
known, the classical Nyquist formula \cite {N28} for a random
electromotive force (emf) $\varepsilon_\omega^2$ has the form:
\begin{eqnarray}
\varepsilon_\omega^2= 2 T ReZ(\omega), \label{E1}
\end{eqnarray}
where T is the temperature in energetic units and $Z(\omega)$ is
the impedance $Z(\omega)$ of a linear electric chain with the
components arranged serially (for example). For a
\emph{quasi-static} field, changing slowly in time, $Z(\omega)$ is
a function of the resistivity $R$, inductance $L$ and capacity $C$
for the electric RLC circuit (the speed of light is taken equal to
unity $c=1$):
\begin{eqnarray}
Z=R-i\left(\omega L -\frac{1}{\omega C} \right)\label{E3}
\end{eqnarray}
In the general case the function $Z(\omega)$ is the
Fourier-component of the linear operator $\tilde{Z}$, which
connects the functions $\varepsilon(t)$ and $J(t)$ in a linear
relation:
\begin{eqnarray}
\varepsilon=\tilde{Z}J
\end{eqnarray}
 or in Fourier-components:
\begin{eqnarray}
\varepsilon_{\omega}=Z(\omega) J_{\omega}.\label{E4}
\end{eqnarray}
For the current fluctuations one derives:
\begin{eqnarray}
\mid J_{\omega}\mid^2= 2T \frac{R'}{R'^{2} +\left(R''+\omega
L-\frac{1}{\omega C}\right)^2}, \label{E5}
\end{eqnarray}
where $R'$ and $R''$ are the real and imaginary parts of the
resistivity $R(\omega)$, respectively. Let us now calculate the
full average square of the magnetic energy of the current
$LJ^2/2$:
\begin{eqnarray}
\frac {L J^2}{2}= \frac{L}{2} \int^{\infty}_{-\infty} \frac
{d\omega}{2\pi} \mid
J_{\omega}\mid^2=LT\int^{\infty}_{-\infty}\frac
{d\omega}{2\pi}\frac{R'}{R'^2+\left(R''+\omega L-\frac{1}{\omega
C}\right)^2}. \label{E6}
\end{eqnarray}

For the case of $\omega$-independent (and, therefore, real)
resistivity $R_0$, which was considered in \cite{KL1},
Eq.~(\ref{E6}) can be written in the form:
\begin{eqnarray}
\frac {L J^2}{2}=T\int^{\infty}_{-\infty}\frac
{d\xi}{2\pi}\frac{1}{1+\left(\xi-\frac{L}{CR_0^2\xi}\right)^2}.
\label{E7}
\end{eqnarray}
in which $\xi=L\omega/R_0$. The integral in Eq.~(\ref{E7}) can be
calculated analytically (or by use of, e.g., the mathematical
program "Maple") and is equal to $1/2$. There is no dependence of
the integral (\ref{E7}) on the parameter $L/CR_0^2$; therefore it
is also independent of the dissipative parameter $R_0$. The
analytical calculation of the integral (\ref{E7}) for the
considered classical case of $\omega$-independent active
resistivity $R_0$ is presented in the Appendix.

\section{Formulation of the problem, the dissipative and physical parameters }

The argumentation \cite{KL1, VSch04} against the quantum
generalization of the Nyquist theorem \cite {N28}
\begin{eqnarray}
\varepsilon_\omega^2= 2R T_\omega;\;\;
T_\omega=\frac{\hbar\omega}{2}\coth\frac{\hbar\omega}{2T}
\label{E8}
\end{eqnarray}
is essentially based on the notion, that the magnetic energy of
the current (see Eq.~(\ref{E6})), as well as other correlation
functions, calculated over the equilibrium distribution, cannot
include the dissipative parameter \cite{KL1, VSch04}. More
concretely, this statement has been formulated as the requirement
for the integral of the magnetic energy $LJ^2/2$ to be independent
of the active resistivity $R_0$. This statement does not apply to
the case of frequency-dependent $T_\omega$, when the magnetic
energy is determined by an equation similar to Eq.~(\ref{E6}). In
the quantum case, however, the integral in (\ref{E6}) contains
$T_\omega$, instead of the temperature $T$ in the classical case,
and depends on $R_0$. According to \cite{KL1, VSch04} it is the
reason to reconsider the quantum Nyquist formula and the other
quantum fluctuation-dissipation relations or theorems (FDT). We
want to stress that in \cite{G51} the independence of the charge
fluctuations in the electric circuit has been considered on the
basis of thermodynamic requirements only for the quasi-static case
($\hbar \omega<T$), e.g., in the classical case.

At first we have to consider and analyze the requirement of
independence of the equilibrium correlation functions of the
"dissipative parameters", as the basic one for reconsideration of
the quantum FDT, in more detail. The simplest way to do that is to
analyze microscopically the systems with Coulomb interaction,
possessing a well-known Hamiltonian. It is evident, that the
"dissipative parameters", such as, e.g., resistivity, themselves
are functions of the fundamental non-dissipative parameters,
namely, the mass $m_\alpha$, the charges $e_\alpha$, the particle
density $n_\alpha$ (for the finite samples the characteristic
sizes can appear) and also the temperature $T$. Therefore it is
necessary to clarify this statement by considering the possible
dependence of the equilibrium correlation functions on the
dimensionless parameters. We can suggest, for example, that for
plasma-like systems the integral over frequencies of the
current-current correlation function $I\equiv J^2$  is of the form
$I = T \Omega_p^2\, f(\Gamma, \lambda, Z)$, where $\Omega_p$, $Z$,
$\Gamma \sim e^2n^{1/3}/T$ and $\lambda \sim \hbar^2n^{2/3}/mT$
are the plasma frequency, ion charge, interaction and degeneration
parameters, respectively (the derivation of this statement is
presented in Section III).

Instead of $\Gamma$ and $\lambda$ it is sometimes more convenient
to use, as independent parameters, e.g., the parameters $\Gamma$
and $\mu = d/a_0 \equiv \Gamma/\lambda$, where $d=n^{-1/3}$ and
$a_0$ is the Bohr radius. In a finite system additional
dimensionless parameters connected with the sizes of the sample
can be present. In some particular case this integral can be
independent of $\Gamma$ and has a form $I \simeq T \Omega_p^2$.

It seems natural to suggest that the statement on independence of
the equilibrium correlation functions of the "dissipative
parameters" in \cite{KL1} can be formulated (in the case of the
Coulomb systems) as the condition $I \simeq T \Omega_p^2$.

Otherwise, if there is dependence on the other parameters
mentioned above (at first on $\Gamma$, for almost classical
plasma) the dissipation manifests itself in the correlation
function. This conclusion follows, in particular, from the formula
for the collision frequency of the Lorentz plasma
$\nu_0=\pi\Omega_p \Gamma^{3/2}L/8 \sqrt 2$ of a weakly non-ideal
plasma with $Z=1$. Here $L$ is the Coulomb logarithm, which for a
weakly non-ideal $(\Gamma<1)$ classical plasmas can be taken (with
logarithmic accuracy) as $L=\ln (1/\Gamma)$. The frequency $\nu$
is, of course, the "dissipative" parameter in the sense of
\cite{KL1}. Therefore, it is natural to suggest that the
dependence on "dissipative" parameters for $J^2$ means in fact the
dependence on $\Gamma$. In any case, it is necessary to suggest
which kind of dependence of the correlation functions, e.g. $J^2$,
on the selected parameters can, in principle, exist. Only on this
basis can be proved that the FDT are rigorously correct or that in
some cases, e.g., for quantum systems an incorrectness exists.

We turn now to the Callen and Welton theorem (CWT) \cite{CW51}.
The CWT, strictly speaking, is formulated on the basis of
statistical consideration of a system with a well- determined
Hamiltonian and, therefore, can be applied to the Coulomb system
with an arbitrary interaction parameter.

How to apply these results to the "macroscopic" electrodynamics,
which are described in terms of macroscopical parameters as
inductance and capacity? The answer is not quite clear, since an
exact determination of correlation functions for these cases is
lacking - as far as we know.

It should be noted that the work (per unit time) produced by emf
$\varepsilon$ under the moving charges in the linear circuit is
equal to the sum of Joule heating, change of the magnetic field
and the energy of electric field in the capacitor \cite{LL2}:
\begin{eqnarray}
\varepsilon
J=RJ^2+\frac{d}{dt}\left(\frac{LJ^2}{2}+\frac{e^2}{2C}\right).
\label{E9}
\end{eqnarray}
This relation leads to the expression (\ref{E3}) for the frequency
dependent impedance. At the same time the applicability of this
representation is limited to the classical and quasi-static
approximations.

\section{Frequency-dependent resistivity}

Nevertheless, let us assume that the representation (\ref{E3}) can
be formally applied also in the quantum case. Let us also assume
that some prohibition exists for the integrals of the correlation
functions (in the case under consideration the integral of type
~(\ref{E6})) to be dependent on the "dissipative" parameters
(although, as was mentioned above, this requirement is not even
clearly formulated). Under this assumption it seems that the
arguments against the CWT in the quantum case in the approximation
considered in \cite{KL1} could be correct, because the $R_0$
dependence of the integral $I$ is absent in the classical case
while it appears in the quantum case.

However, from our point of view, the arguments against the quantum
generalization of CWT are not correct, because the approximation
used in \cite{KL1} is not general. \emph{The static approximation
for resistivity (and, in general, for the impedance) is not
correct for high frequencies, which are essential in the integral
in Eq.~(\ref{E6})}. The dependence $R(\omega)$ can lead, as we
will show below, to a dependence of the integral on the effective
frequency of collisions if $L\neq0$ and $C\neq\infty$, even in the
classical case.

\subsection{The classical case for $L=0$, $C=\infty$.}

For the special case $L=0$ and $C=\infty$ we will show below,
that: i) the case of $R$ being independent of $\omega$ is
meaningless, because in this limit $J^2\sim1/L\rightarrow\infty$,
whereas it must be finite, ii) if we use $R(\omega)$ we arrive at
CWT for arbitrary interaction between the particles in both
classical and quantum cases, iii) there is no fundamental
prohibition for correlation functions, e.g. the integral $I$, to
be dependent on the "dissipative parameters" (in the sense
discussed above), although in the classical case this integral in
fact is independent of it.

Now we turn to the main reasons showing that CWT is correct. It
was already noted  \cite{Tr04} that the impedance in quasi-static
form, used above, is valid only for very low frequency (in the
classical state as well as in the quantum case). In particular,
for the frequencies $\hbar \omega/ T \geq1$, where the quantum CWT
is applicable, the quasi-static form of the impedance cannot be
used.

We will now show that even for the classical case the quasi-static
approximation for the impedance $Z(\omega)$, described by
Eq.~(\ref{E3}) cannot be used and, therefore, the Nyquist formula
is only an approximation, which is valid only for the quasi-static
case of the small frequencies. At the same time the CWT is
applicable for all frequencies and in this sense is a rigorous
result for classical and quantum systems.

For this purpose we consider the specific circuit with $L=0$ and
$C=\infty$ (capacitor with the distance between the plates $d=0$).
Then, in the quasi-static approximation used in \cite{KL1}, $Z$ is
equal to $R$ in this case and the integral is infinite if $R$ is
$\omega$-independent.

In reality, for this type of circuit (the conductor with $L=0$ and
$C=\infty$) we have:
\begin{eqnarray}
\frac{1}{R(\omega)}=\frac{\sigma(\omega)S}{l}, \label{E10}
\end{eqnarray}
where $\sigma(\omega)$ is the dynamic conductivity, $l$ and $S$
are the length and the area of the conductor section,
respectively. Eq.~(\ref{E10}) is the consequence of the relations
$j(\omega)=\sigma(\omega)E(\omega)$ and $J=Sj$,
$\varepsilon(\omega)=lE(\omega)$ for the linear conductor.

For the simplest case under consideration the approximate
expression for the conductivity can be written in the
Drude-Lorentz form:
\begin{eqnarray}
\sigma_{DL}(\omega)=\frac{ine^2}{m(\omega+i\nu)}, \label{E11}
\end{eqnarray}
where $\nu$ is the effective, $\omega$-independent, collision
frequency of the particles, which transfer the current.
Substituting Eqs.~(\ref{E10}),(\ref{E11}) in Eq.~(\ref{E7}) with
$L=0$ and $C=\infty$ we arrive at the integral:
\begin{eqnarray}
J^2= \int^{\infty}_{-\infty} \frac {d\omega}{2\pi} \mid
J_{\omega}\mid^2=2 DT\int^{\infty}_{-\infty}\frac
{d\omega}{2\pi}\,\mbox{Re}\sigma(\omega). \label{E12}
\end{eqnarray}
For the Drude-Lorentz conductivity we find from Eq.~(\ref{E12})
\begin{eqnarray}
J^2= 2 DT\int^{\infty}_{-\infty}\frac
{d\omega}{2\pi}\,\mbox{Re}\sigma_{DL}(\omega)=\frac{2DTne^2}{m}\int^{\infty}_{-\infty}\frac
{d\omega}{2\pi}\,\frac{\nu}{\omega^2+\nu^2}. \label{E12a}
\end{eqnarray}
The result of the integration is, naturally, finite and equals
$DTne^2/m\equiv DT \Omega_{pe}^2/4\pi$, where $D\equiv S/l$.
Apparently, this result does not involve the dissipative parameter
$\nu$. This implies that $J^2$ does not involve the interaction
parameter $\Gamma$ (we can consider this fact as a specific
condition for the non-dissipative character of correlation
functions, see above).

It is necessary to stress, that in Eq.~(\ref{E12}) we used the
so-called internal conductivity (\ref{E11}), which we choose in
the Drude-Lorentz approximation. This conductivity is the response
function for the self-consistent electrical field in Coulomb
systems. The general relation between the internal conductivity
$\sigma(\omega)$ and the external conductivity
$\sigma_{ex}(\omega)$ (the response on an external field applied
to the system) has the form \cite{AP77}:
\begin{eqnarray}
\sigma(\omega)= \frac{\sigma_{ex}(\omega)}{1-\frac{4\pi i
\sigma_{ex}(\omega)}{\omega}},\label{E12b}
\end{eqnarray}
It is easy to show by straight calculation that the integral
(\ref{E12}) has the same value for $\sigma(\omega)$ and
$\sigma_{ex}(\omega)$ if we use the Drude-Lorentz approximation.
In the general case for the exact (non-approximative) expressions
for conductivities it is also true \cite{AT03}, as can be shown by
using of the Kramers-Kronig relations for $\sigma_{ex}(\omega)$
and $\sigma(\omega)$. The problem of the validity of the
Kramers-Kronig relations for conductivities has been investigated
by Kirzhnits in \cite{Kr76}. In the long-wavelength limit not only
$\sigma_{ex}$, but also the dielectric function
$\varepsilon(\omega)$ and therefore $\sigma(\omega)$ satisfy the
Kramers-Kronig relations (w.r.t. the external conductivity this
statement is valid also for finite values of the wave vector $k$).
The resistivity, present in the relations (\ref{E3}),(\ref{E12}),
is the internal one, because only the internal conductivity
$\sigma$ has a finite limit for $\omega\rightarrow0$ and provides
a finite value of $R'(\omega\rightarrow0)\equiv
ReR(\omega\rightarrow0)$. The external coductivity in the limit
$\omega\rightarrow0$ tends to zero as $\omega^2$ (see, e.g.,
\cite{PN66}).

By applying the sum rule \cite{Ku57} for the external conductivity
of a system of charges with arbitrary strong interaction between
the particles:
\begin{eqnarray}
\int^{\infty}_{-\infty}\frac
{d\omega}{\pi}\,\mbox{Re}\sigma_{ex}(\omega)=\sum_{r}\frac{e_r^2
n_r}{m_r} \label{E13}
\end{eqnarray}
with $r$ the index of the species of particles, we arrive at a
general relation similar to Eq.~(\ref{E12}), but without any
assumption about the concrete form of the conductivity for the
classical system:
\begin{eqnarray}
J^2= DT\int^{\infty}_{-\infty}\frac
{d\omega}{\pi}\,\mbox{Re}\sigma_{ex}(\omega)=DT
\sum_{r}\frac{e_r^2 n_r}{m_r}. \label{E14}
\end{eqnarray}
The additional assumption that has been adapted to establish
Eq.~(\ref{E13}) is isotropy of the system under consideration.
This means, in particular, that the conductivity tensor is
diagonal with equal components. According to the statements
considered above, Eqs.(\ref{E13}),(\ref{E14}) are also satisfied
if instead of $\sigma_{ex}$ we write $\sigma$.

As is evident from the general point of view, the current-current
correlation function in the classical case is independent of the
"dissipative" parameters.

\subsection{The case of a classical system with $L\neq0$, $C\neq\infty$.}

Let us now consider, as an example of the classical case, the
integral (\ref{E6}) with $L\neq0$ and $C\neq\infty$. We assume
that the terms related with the inductance and capacitor in
Eq.~(\ref{E3}) for the impedance are unchanged. For the
Drude-Lorentz internal resistivity with constant $\nu(\omega)=\nu$
the real and imaginary parts of the resistivity are equal to $Re
R_{in}(\omega)=R_0 \equiv m \nu/D n e^2$ and $Im R_{in}
(\omega)=-m\omega/D ne^2$, respectively, where $Re R_{in}
(\omega)$ is independent of $\omega$. For the external resistivity
in Drude-Lorentz approximation we obtain $Re R (\omega)=R_0 \equiv
m \nu/D n e^2$ and $Im R (\omega)=-m(\omega^2-\Omega_p^2)/D \omega
ne^2=-R_0 (\omega^2-\Omega_p^2)/\nu\omega$.

Then we return to Eq.~(\ref{E7}) and to the equality
$LJ^2=TL/\tilde L$ with the effective $\tilde L = L-(m/D ne^2)$.
It is evident that the condition $\tilde L>0$ must be fulfilled,
which demonstrates the limited applicability of the above
formulated assumption. However, if we take into account that the
expression (\ref{E13}) is the result for the thermodynamic limit,
whereas the Nyquist formula is written for finite samples, we have
to consider large values of $D$ and on this basis we can suggest
$\tilde L>0$.

In general, we have to stress the difference between FD relations
for finite systems and for systems in the thermodynamic limit.
Consideration of the case $L\neq0$ and $C\neq\infty$ has
conventional character, because the value $\tilde L$ can change
sign, although the value $J^2$ must be positive according to
definition. This means that representation (\ref{E6}) has,
strictly speaking, to be additionally modified for the case
$L\neq0$ and $C\neq\infty$, when we take into account the
$\omega$-dependence for the resistivity $R$.

Nevertheless, it is easy to see, even for the restricted
modification of Eq.~(\ref{E6}), when we use the Drude-Lorentz
conductivity with $\omega$-dependent $\nu(\omega)$ the resistivity
$R_0\rightarrow R_0 \theta (\omega)$ (in general, $\theta(\omega)$
has real and imaginary parts $\theta'(\omega)$ and
$\theta''(\omega)$) and the statement about the energy $LJ^2$
being independent of dissipative parameter is not valid. To show
this, let us consider the conductivity of the Lorentz classical
plasma system:
\begin{eqnarray}
\sigma(\omega)=\frac{ie^2}{3T}\int \frac{v^2
f_0(p)}{\omega+i\nu(v)}\, d^3p, \label{E7aa}
\end{eqnarray}
where $f_0(p)$ is the Maxwellian distribution, normalized to the
electron density $n_e$, and $\nu(v)=4\pi Z e^4 n_e L/m^2 v^3$ is
the velocity-dependent effective collision frequency for charged
particles. In this approach we use the notation $R_0\equiv
m\nu_0/Dne^2$. The functions $\theta'(\omega)$ and
$\theta''(\omega)$ can be found by using the relation
$R=1/D\sigma$, where $\sigma$ is determined by (\ref{E7aa}). As
follows on the basis of Eq.~(\ref{E7aa}), the function
$\theta'\neq const$ and $\theta''$ is a nonlinear function of
$\omega/\nu_0$. In particular, in the high-frequency region
($\omega>\nu_0$) the functions $\theta'(\omega)$ and
$\theta''(\omega)$ are equal to:
\begin{eqnarray}
\theta'(\omega)=\frac{32}{3\pi} \frac{\omega^2}{\omega^2+\nu_1^2},
\label{E7ab}
\end{eqnarray}
\begin{eqnarray}
\theta''(\omega)=-\frac{\omega^3}{\nu_0(\omega^2+\nu_1^2)},
\label{E7ac}
\end{eqnarray}
where $\nu_1=32\nu_0/3\pi$. This means that the integral
(\ref{E6}) cannot be represented in a form similar to (\ref{E7})
with some effective constants $L$ and $C$ and depends on the
"dissipative" parameter $R_0$.

Therefore in the case of a system with $L\neq0$ and $C\neq\infty$
the statement about independence of the correlation function of
"dissipative parameters" is not true even in the classical case,
at least for a consideration based on the same dependence
$Z(\omega)$ on $L$ and $C$ as in Eq.~(\ref{E3}).

\subsection{The quantum system with $L=0$ and $C=\infty$, Callen-Welton relation.}

We now turn to the quantum case with $L=0$ and $C=\infty$. The sum
rule ~(\ref{E13}) is valid in the quantum case as well
\cite{Ku57}. At the same time the correlation function $J^2$ has
to be written (see, e.g. \cite{Ku57}) as:
\begin{eqnarray}
J^2= D\int^{\infty}_{-\infty}\frac {d\omega}{\pi}\,T_\omega
\mbox{Re}\sigma(\omega). \label{E15}
\end{eqnarray}

This result can be easily obtained from the general exact CWT for
the density current correlation function:
\begin{eqnarray}
<j_i({\bf r},t)j_k({\bf r'},t')>=\frac{\hbar
\omega}{4\pi}\coth\left(\frac{\hbar
\omega}{2T}\right)\left[\sigma_{ik}(\omega)+\sigma_{ki}^\ast(\omega)\right]\delta({\bf
r}-{\bf r'}) \label{E15a}
\end{eqnarray}
where $\sigma_{ik}$ is a component of the tensorial conductivity
and the asterisk denotes the complex conjugate by integration over
${\bf r}$ and ${\bf r'}$ in the volume of a linear conductor
$V=lS$ and the suggestion that in this volume the conductivity is
the same as in the thermodynamic limit. Under this suggestion the
result coincides with the Nyquist relation in the form
~(\ref{E15}), where conductivity is an exact function of the
frequency $\omega$.

There exist \emph{no arguments} for independence of this integral
of the parameter $\Gamma$. Therefore this integral depends on
"dissipative parameters". Let us demonstrate this fact in the
example of the expansion of $J^2$ in the parameter $\hbar
\omega/T$. In this case the Eq.~(\ref{E15}) can be rewritten
approximately in terms of the zeroth and second momenta of the
conductivity:
\begin{eqnarray}
J^2 \simeq D\int^{\infty}_{-\infty}\frac {d\omega}{\pi}\,T_\omega
\mbox{Re}\sigma(\omega)\simeq \frac
{DT}{\pi}(\mu_0+\frac{\hbar^2\mu_2}{12T^2 }), \label{E16}
\end{eqnarray}
where
\begin{eqnarray}
\mu_{2n}\equiv\int^{\infty}_{-\infty} d\omega \omega^{2n}
Re\sigma(\omega). \label{E17}
\end{eqnarray}
The term with $\mu_0=\Omega_p^2/4$ leads to the classical value of
$J^2$ (see Eq.~(\ref{E15})). The second moment of the external
conductivity has been calculated in \cite{AT03} and can be
expressed in terms of the value of the electron-ion equilibrium
correlation function $g_{ei}(\textbf{r})$ at $r=0$. It can be
connected with the second moment of the internal conductivity by
use of the asymptotic expansions for the Kramers-Kronig formulas.
For the case $Z=1$ the second moment of the internal conductivity
equals:
\begin{eqnarray}
\mu_{2}= \frac {\Omega_{pe}^4 g_{ei}(0)}{24}. \label{E18}
\end{eqnarray}
The final expression for $J^2$ following from the relations
mentioned above is:
\begin{eqnarray}
J^2 \simeq \frac
{DT\Omega_{pe}^2}{4\pi}\left(1+\frac{\hbar^2\Omega_{pe}^2g_{ei}(0)}{6T^2
}\right), \label{E19}
\end{eqnarray}

Naturally the function $g_{ei}(\textbf{r})$, as well as it value
in the point $r=0$, is a function of the parameter $\Gamma$. To
calculate $g_{ei}(0)$ for a purely Coulomb system the interaction
and quantum effects in $g_{ei}$ also have to be taken into
account, in particular to avoid the divergence $g_{ei}(0)$ for
small distances. For example, for a weak plasma interaction
$\Gamma\ll1$ the pair correlation function can be written (Z=1) as
\cite{KKER86}:
\begin{eqnarray}
g_{ei}(r)= S^{(2)}_{ei}(r) \exp\left[\frac{e^2}{4 \pi T
r}(1-e^{-\kappa r})\right], \label{E20}
\end{eqnarray}
where $\kappa^{-1}=\sqrt {T/4\pi e^2n}$ is the Debye radius (with
the density $n=n_e+n_i$) and $S^{(2)}_{ei}(r)$ is the pair
electron-ion Slater sum, which takes into account the quantum
effects. The value $g_{ei}(0)$ for the approximation
(\ref{E15}))equals:
\begin{eqnarray}
g_{ei}(0)=
S^{(2)}_{ei}(0)\exp\left(\frac{\Gamma^{3/2}}{\sqrt{4\pi}}\right),
\label{E21}
\end{eqnarray}
The Slater sum $S^{(2)}_{ei}(0)$ is a function of the parameter
$\zeta=e^2/4\pi T \lambda_{ei}$, where
$\lambda_{ei}=\hbar/\sqrt{2mT}$ and $m$ are De Broglie wavelength
and the reduced electro-ion mass respectively. This function has a
finite value \cite{KKER86} for all values of the parameter
$\lambda_{ei}$, except $\lambda_{ei}=0$.

Therefore $J^2$, even in the simple approximations used above,
depends on the "dissipative" parameter $\Gamma$. The dependence of
the current-current equilibrium correlation functions on the
"dissipative" parameters is not in contradiction with any
fundamental principles of quantum statistical theory.

\section{Conclusions.}

The main conclusion of the present paper is the statement that the
quasi-static approximation for the impedance in the form used in
\cite{KL1, N28} is not allowed when considering the opportunity
for violation of the general quantum CWT. Therefore, presently
there are no arguments to discuss violation of the CWT in the
quantum case on this basis. As it is, strictly speaking, the CWT
is proved only for the case of a conductor with $L=0$ and
$C=\infty$ and for an infinite medium, when the Hamiltonian
description of the system is well-determined. For this case we
have shown the way to connect the macroscopical model of the
Nyquist fluctuation-dissipation relation with the microscopic
derivation of the CWT by introduction of the frequency dependent
resistivity in the Nyquist model. We also demonstrated the absence
at this time of a clear way for generalization of this relation
for the case $L\neq0$ and $C\neq\infty$, in spite of some attempts
to do that \cite{VSch04}. The important question about
determination of the "dissipative" parameters has been discussed
above for plasma-like systems and connected, in particular cases
considered above, with the $\Gamma$-dependence of the
current-current correlation function. In general, a more
fundamental determination of the dissipative parameters has to be
formulated.

In the framework of the kinetic equations, the dissipative
parameters are connected with the respective effective collision
frequencies. For the problem of FD relations the approach of
kinetic equations is not enough, because the results have to be
found for arbitrarily strong interaction between the particles. In
this case the determination of the dissipative parameters is
based, in general, on the imaginary parts of the poles of the
analytical continuation of the temperature Green functions
\cite{AGD62}. In the particular cases of weak or short-range
interactions between the particles these poles lead to the same
results as the kinetic equations approach. The problem of the
validity of summation of the rows of the perturbation theory for
the Green functions, as well as the problem of irreversible
behavior of physical systems due to damping, connected with these
poles, have a fundamental character and still have not been
completely clarified. Nevertheless, we trust that the general
results, connected with dissipation based on the Green function
approach are applicable to obtain the fluctuation-dissipation
relations for systems with well-defined Hamiltonian. Of cause the
$\Gamma$- dependence of the current-current correlation function,
which we consider as a manifestation of the dependence on
"dissipative parameters", can be related in fact not only with a
shifting of the imaginary part of the poles of the Green
functions, but also with a shifting of the real parts of these
poles. Nevertheless we do not see the opportunity to distinguish
in the exact final result for CWT the "dissipative" and
"non-dissipative" dependence on the parameter $\Gamma$.

The sense of the consideration, suggested by Yu.L. Klimontovich
\cite{KL1} is, from our point of view, not about incorrectness of
the CWT, but about limited applicability of the Nyquist formula in
the approximation of constant $R$. This limitation follows, at
first, as was shown above, from the quasi-static character of the
impedance in the form (\ref{E3}). It is evident not only in the
quantum, but also in the classical case. At the moment there is no
clear basis for a general and rigorous consideration of the
electrical circuit with finite $L\neq0$ and $C\neq\infty$ in the
quantum and even in the classical case, except the model of
quasi-static approximation for the impedance, which
is applicable nevertheless to many experimental situations.\\

\section{Appendix. }

The integral $I$ in Eq.~(\ref{E7}),
\begin{eqnarray}
I= \frac {1}{\pi}\int^{\infty}_{-\infty}
d\xi\frac{1}{1+\left(\xi-\frac{L}{CR_0^2\xi}\right)^2}. \label{A1}
\end{eqnarray}
can be calculated analytically. According to the Maple calculation
program this integral equals one. We can rewrite the integral
(\ref{A1}) in the form
\begin{eqnarray}
I= \frac {1}{\pi}\int^{\infty}_{-\infty}
d\xi\frac{\xi^2}{\xi^2+\left(\xi^2-\lambda\right)^2}. \label{A2}
\end{eqnarray}
with $\lambda\equiv L/CR_0^2>0$. The denominator is equal to
$(\xi^2-\xi_1^2)(\xi^2-\xi_2^2)$, where the roots are:
\begin{eqnarray}
\xi_1^2=\frac{1}{2}\left[-(1-2\lambda)+\sqrt{(1-4\lambda)}\right],\;
\xi_2^2=\frac{1}{2}\left[-(1-2\lambda)-\sqrt{(1-4\lambda)}\right]
\label{A3}
\end{eqnarray}
The integral (\ref{E7}) can be split into two integrals
$I=I_1+I_2$, with
\begin{eqnarray}
I_1=\frac{1}{\pi\sqrt{(1-4\lambda)}}\int^{\infty}_{-\infty} d \xi
\frac{\xi_1^2}{(\xi^2-\xi_1^2)}, \;
I_2=-\frac{1}{\pi\sqrt{(1-4\lambda)}}\int^{\infty}_{-\infty} d \xi
\frac{\xi_2^2}{(\xi^2-\xi_2^2)}\label{A4}
\end{eqnarray}
Let us consider the particular case $1-4\lambda>0$. It is easy to
see that the roots $\xi_1^2, \xi_2^2$ in this case are real and
negative. Therefore the integrals are equal to:
\begin{eqnarray}
I_1=-\frac{\sqrt{\mid\xi_1^2\mid}}{\sqrt{(1-4\lambda)}}, \; I_2=
\frac{\sqrt{\mid\xi_2^2\mid}}{\sqrt{(1-4\lambda)}}. \label{A5}
\end{eqnarray}
Because $\mid\xi_2^2\mid >\mid\xi_1^2\mid$ the sum $I=I_1+I_2>0$
and equals unity. To show this it suffices to take $I^2$ and to
use Eq.~(\ref{A5}). The same is true for an arbitrary value of
$\lambda$.

\section*{Acknowledgment}

Authors are gateful to W. Ebeling, A.M.\,Ignatov, N.I.
Klyuchnikov, and I.M. Tkachenko for valuable
discussions of the various problems, reflected in this work,
which has been submitted in Physica A (8 March 2006).\\

\end{document}